\documentclass[twocolumn,prb,aps,amsmath,amssymb,floatfix]{revtex4-1}

\usepackage{amsmath}    
\usepackage{graphicx}   
\usepackage{verbatim}   
\usepackage{color}      
\usepackage{subfigure}  
\usepackage{hyperref}   

\begin{document}

\title{Thermoelectricity of the Ferromagnetic Superconductor UCoGe}
\author{Liam Malone$^1$, Ludovic Howald$^2$, Alexandre Pourret$^2$, Dai Aoki$^2$, Valentin Taufour$^2$, Georg Knebel$^2$ and Jacques Flouquet$^2$}

\affiliation{$^1$Laboratoire National des Champs Magnétiques Intenses, UPR 3228 (CNRS-INSA-UJF-UPS), Toulouse 31400, France \\ $^2$ INAC, SPSMS, CEA Grenoble, 17 Rue des Martyrs, 38054 Grenoble, France }

\begin{abstract}
UCoGe  exhibits superconductivity in the presence of ferromagnetism.  When a field is applied along the \textbf{b} axis (perpendicular to the easy axis), ferromagnetism is weakened and superconductivity is enhanced.  This enhancement has been attributed to an increase in coupling as observed in the enhanced effective mass produced by the critical fluctuations as the ferromagnetic transition is strongly suppressed. However it is also important to know if and how the Fermi surface changes near the critical point.  Here we report measurements of the thermoelectricity of UCoGe which reveal a low carrier density metal.  Under magnetic field applied along the \textbf{b} axis, a sharp peak is observed in the thermopower of UCoGe at H$^*=11.1$\,T and low temperature which becomes broader at higher temperatures. At higher field, the thermopower changes sign which suggests a modification of the Fermi Surface. We analyze these results using a topological change in Fermi surface and show that this can explain both the thermopower and the enhanced superconductivity.

\end{abstract}
\maketitle

\section{Introduction}

The discovery of superconductivity coexisting with ferromagnetism in three uranium based compounds UGe$_2$ \cite{Saxena2000}, URhGe \cite{Aoki2001} and most recently UCoGe \cite{Huy2007} has attracted much interest due the unusual properties of the superconducting state and the proximity to a critical pressure where the ground state of each compound switches from ferromagnetic (FM) to paramagnetic (PM).  The enhanced ferromagnetic fluctuations in these Ising ferromagnets give rise to unconventional phenomena such as reentrant superconductivity \cite{Aoki2009} \cite{Levy2005} and spin triplet superconductivity \cite{Huy2008}, \cite{Slooten2009}.  In the cases of URhGe and UCoGe, a moderate field applied perpendicular to the easy axis can strongly suppress the Curie temperature to zero temperature \cite{Levy2007}, \cite{Aoki2009}. For URhGe this reveals a reentrant superconducting phase and a reorientation of the magnetic moment \cite{Levy2005} to be parallel to the applied field.  For UCoGe, an enhancement of the superconducting temperature is observed \cite{Aoki2009}.

The phase diagram of a weak itinerant ferromagnet has been the subject of recent theoretical \cite{Belitz2005} and experimental work \cite{Taufour2010}, \cite{Kotegawa}.  Measurements on UGe$_2$ under pressure ($P$) and field ($H$) \cite{Taufour2010} (applied along the easy axis) have demonstrated that the phase diagram has a tricritical point where the transition from the paramagnetic state (PM) to the ferromagnetic state (FM) becomes first order and bifurcates in the ($H-P$) plane resulting in quantum critical end points at high field and high pressure \cite{Taufour2010} which is in rough agreement with the theoretical predictions \cite{Belitz2005}.  However the changes in Fermi surface in UGe$_2$ \cite{Haga2002} are not accounted for in the theory even though these changes can explain some of the features of UGe$_2$ \cite{Sandeman2003}. Also it has been shown that topological transitions near quantum critical points can lead to unconventional critical behavior \cite{Yamaji2007}.

In this paper, we focus on UCoGe. At ambient pressure, this compound is a weak ferromagnet with a Curie temperature $T_{C}\sim$2.8\,K associated with a spontaneous moment of $M_0\sim$0.04\,$\mu_B$ directed along the \textbf{c} axis at $T$=0\,K \cite{Huy2007}.  The superconducting transition temperature $T_{SC}\sim0.6$\,K in zero field.  Applying pressure suppresses ferromagnetism and enhances superconductivity until T$_{C}\sim T_{SC}$ at around 1.25\,GPa \cite{Hassinger2008}.  Above this pressure, the ferromagnetic transition is not observed.  This pressure dependence could imply that UCoGe sits very close to the tricritical point. NMR measurements \cite{Ohta2010} have suggested that the PM-FM transition is first order in UCoGe. Applying a field along the \textbf{b} axis also suppresses ferromagnetism \cite{Aoki2011b}\cite{Hardy2011} and enhances superconductivity resulting in a unconventional 'S-shaped' upper critical field H$_{c2}$ curve \cite{Aoki2009}.

Thermoelectric effects are very sensitive to changes in the Fermi surface topology and effective mass enhancements as has been in shown in several studies, for example on CeRu$_2$Si$_2$ \cite{Amato1988} and CeColn$_5$ \cite{Izawa2007}. Quantum oscillation experiments \cite{Aoki2010} on UCoGe have revealed an unusual field dependence of the cyclotron masses in fields greater than 20\,T along the \textbf{b} axis. In this paper we report measurements of the thermopower of UCoGe as a function of field applied along the \textbf{b} axis to examine these effects in the region of reentrant superconductivity ($H_b\sim 10$\,T).  We suggest that the Fermi surface of UCoGe undergoes a topological change as the ground state switches from FM to PM under applied field and compare this with recent band structure calculations.  We also argue that this change in Fermi surface can be used to explain the unusual $H_{c2}$ curve.

\section{Experimental Method}

Single crystals of UCoGe were grown using the Czhochralski method in a tetra-arc furnace. The residual resistivity ratio (RRR) of the measured sample was around 30. The thermopower was measured using the two thermometer, one heater technique at temperatures down to 100\,mK and in fields up to 18\,T.  All measurements were performed in both positive and negative fields and averaged to prevent contamination of the Nernst signal in the Seebeck.  Thermometers were calibrated against a Germanium thermometer in a field compensated region up to 16\,T and down to 100\,mK.  The temperature gradient was applied along the \textbf{a} axis of the crystal and the field applied along the \textbf{b} axis in all measurements.  Based on the angular dependance of the upper critical field $H_{c2}$ curves \cite{Aoki2009} the crystals were aligned along the \textbf{b} axis to $<5^{\circ}$ with respect to the magnetic field $H$.  The setup also allows in-situ measurement of the resistivity therefore all data presented below was taken on the same sample in the same conditions.

\section{Results}

\begin{figure}\label{Fig1}
\includegraphics[width=70mm]{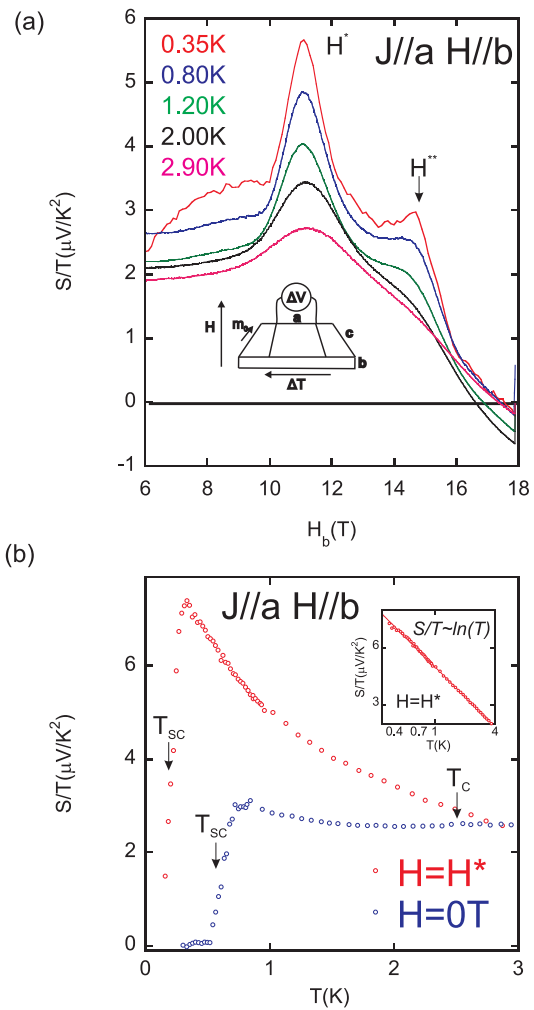}
\caption{(Colour Online)(a) Thermopower divided by temperature $S/T$ as a function of field applied along the \textbf{b} axis $H_b$ at several temperatures.  A large peak is observed at $H^*=11.1$\,T at all temperatures, a smaller anomaly is observed at $H^{**}=14.6$\,T at low temperature and a change of sign is observed above 16\,T.   Inset is a diagram of the setup showing the applied field $H$, the temperature gradient $\Delta T$, the measured thermoelectric voltage $\Delta V$ and the zero field magnetization $M_0$ compared with the crystal axis. (b) Temperature dependence at two fields 0\,T and $H^*$.  In 0\,T both the Curie temperature $T_c$ and the superconducting transition $T_{SC}$ are observed.  For $H=H^*$ a possible logarithmic divergence is observed above $T_{SC}$. Inset shows $H=H^*$ data on a semilog plot illustrating the divergence above $T_{SC}$. }
\end{figure}

Figure 1 shows the temperature and field dependence of the thermopower divided by temperature $S/T$ of UCoGe.  In 0\,T, $S/T$ shows a small anomaly at $T_C$=2.8\,K which is followed by a slight increase of $S/T$ until the superconducting transition $T_{SC}$=0.6\,K.  In a free electron gas, $S/T$ is constant at low temperature and it's amplitude is inversely proportional to the Fermi temperature with a sign which reflects the type of charge carrier.  For UCoGe, $S/T$ is $\sim 3\mu$ V/K$^{2}$ just above $T_{SC}$ indicating the dominant carriers are hole like.  In a free electron gas, the ratio $q$($=\frac{S}{T}\frac{N_ae}{\gamma}$, $N_a$ is Avogadro's number) of the thermoelectric power to the Sommerfeld coefficient $\gamma$ of the linear term in specific heat is a constant \cite{Behnia2004}.  It has been observed in a large variety of strongly correlated systems that the low temperature value of $S/T$ scales with the Sommerfeld coefficient via ratio $q$ which is inversely proportional to carrier number. Further more despite the complexity of multiband sytems it has been shown \cite{zlatic2007} that even in compensated cases (such as URu$_2$Si$_2$) the thermoelectric response is dominated by one carrier type and thus the scaling factor $q$ holds. For UCoGe, $q=5$ ($\gamma=57$\,mJ/molK$^2$ \cite{Huy2007}) which indicates UCoGe has a low carrier density. This is similar to the case of URu$_2$Si$_2$ which also has $q\sim5$ and is a low carrier semi-metal.  Recent quantum oscillation measurements \cite{Hassinger2008}, \cite{Altarawneh2011} and thermopower measurements \cite{Malone2011} imply the Fermi surface of URu$_2$Si$_2$ is strongly modified in a moderate field (of the order of 20\,T applied along the easy magnetization axis). This could be the same in UCoGe as the quasiparticle masses are heavy \cite{Georg}, \cite{Aoki2010} and this coupled with the low carrier density leads to a small bandwidth which can be strongly modified by a moderate field.

\begin{figure*}\label{Fig2}
\includegraphics[width=160mm]{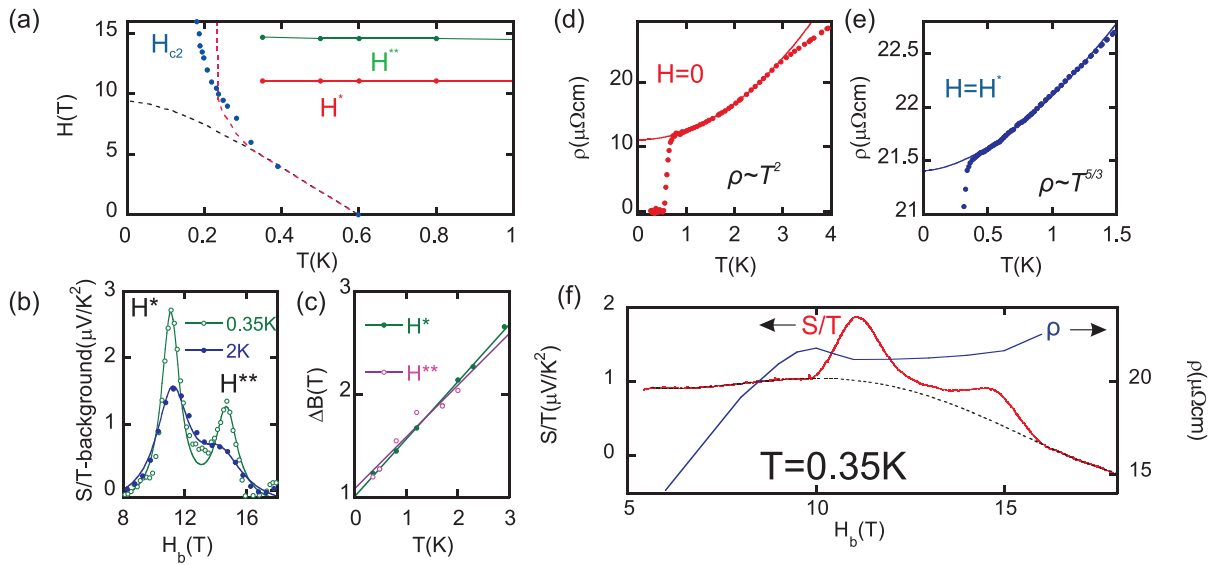}
\caption{(Colour Online)(a) Phase diagram of UCoGe in the $(T-H_b)$ plane using thermopower data and resistivity data showing $H_{c2}$, $H^*$ and $H^{**}$.  The dashed lines are a calculation for a polar p-wave state (black) and the same calculation assuming a variation of $v_F$ with field (pink). (b) $H^*$ and $H^{**}$ at two different temperatures with background subtracted.  Lines are a fit to two Lorentz distributions, one centered on $H^{**}$ and one on $H^*$ (some points not shown for clarity). (c) Width of $H^*$ and $H^{**}$ taken from the Lorentz distribution fit as a function of temperature.  The lines are linear fits. (d) Resistivity data at 0T as a function of temperature.  The line is fit to $\rho=\rho_0+AT^2$ below $T_C$. (e) Resistivity data at $H^*$ as a function of temperature.  The line is fit to $\rho=\rho_0+aT^{\frac{5}{3}}$ below 1.5\,K. (f) Field dependence of thermopower and resistivity taken at 0.35\,K.  Dashed line is the assumed background used for Lorentzian fits in (b).}
\end{figure*}

As the field is increased, $S/T$ remains roughly constant ($\sim$3\,$\mu$V/K$^2$) up to $\sim$10\,T.  At all measured temperatures (up to 3.5\,K), a peak is observed centered on $H^*$=11.1\,T.  This field is independent of temperature but at higher temperatures the peak is broader and less pronounced.  The peak is present even above the zero field $T_{C}$.   Above $H^*$, at low temperature a second anomaly is observed at $H^{**}$=14.6\,T which is smaller and indistinguishable from the background at temperatures above $\sim2$\,K.  At higher field a change of sign of $S/T$ is observed indicating a change in the dominant carrier type and therefore a Fermi surface change.  This is the field range where quantum oscillations have been observed \cite{Aoki2010}.  There is a strong temperature dependence of $S/T$ at $H^*$, with the magnitude of $S/T$ increasing up to $\sim$7\,$\mu$V/K$^2$ at the superconducting transition temperature.  This could indicate one of two possibilities:  $H^*$ could be a critical point where one would expect an enhancement of the effective mass and hence thermopower \cite{Hartmann2010}, \cite{Izawa2007} or $H^*$ could be a topological change in the Fermi surface which can give a large thermopower anomaly at low temperature as described by Lifshitz \cite{Lifshitz1960}.  The fact that $H^*$ is visible above the Curie temperature is interesting.  If it is a Fermi surface instability then it implies that an element of the Fermi surface is constant in the ferromagnetic and paramagnetic state at low fields which is contrary to recent band structure calculations \cite{Samsel-Czekaa2010} but consistent with fact that the zero field thermopower shows only a small anomaly at $T_{C}$.  If it is a critical point then the fluctuations are felt to high temperature perhaps indicating a quantum critical point at $H^*$.  It should be noted that the thermopower data implies there is a topological change between the zero field PM state and the field induced PM state as the change of sign occurs even at temperatures above $T_C$.  This could be evidence that the comparable scale of the Fermi energy and Zeeman energy is at least partly responsible for the modification of the Fermi surface such as is observed in URu$_2$Si$_2$ \cite{Altarawneh2011}, \cite{Malone2011}. Finally strong evidence for a Lifshitz transition is that the maximum thermoelectric response at $H^*$ is enhanced compared with the response expected from the effective mass enhancements measured in a perfectly aligned crystal \cite{Aoki2009}.

Figure 2 (a) shows the superconducting phase diagram of UCoGe derived from thermopower and resistivity data.  The so called 'S-shaped' $H_{c2}$ curve is slightly suppressed probably due to the slight sample misalignment and is visible as a sudden upturn in the $H_{c2}$ curve.  The calculated $H_{c2}$ curve for a polar p-wave state is also shown calculated in \cite{Scharnberg1980} and following \cite{Huy2008} based on the initial slope of $H_{c2}$.   $H^*$ seems to coincide with the near vertical region of $H_{c2}$ suggesting that it  drives the enhanced superconductivity phase. Figure 2(b) shows the thermopower in the critical region after a smooth background is subtracted.  The two peak structure can be reasonably well fitted with two Lorentzians centered on H$^*$ and H$^{**}$. The fit is to the function $S=S^{**}\left(\frac{\Delta H^{**2}}{\left(H-H^{**}\right)^2+\Delta H^{**2}}\right)+S^*\left(\frac{\Delta H^{*2}}{\left(H-H^*\right)^2+\Delta H^{*2}}\right)$ where $S^{**}$ ($S^*$) is the amplitude of the peak at $H^{**}$ ($H^*$) and $\Delta H^{**}$ ($\Delta H^*$) is the width of the peak at $H^{**}$ ($H^*$).  Both peaks are well fitted with a similar width but different amplitudes.  Figure 2(c) shows $\Delta H$ as a function of temperature for both peaks and is linear in temperature with a finite intercept.  This is consistent with a change in Fermi surface as temperature broadens the effect but the field required for the change is constant.  As the temperature dependence of the width of the $H^*$ and $H^{**}$ anomalies are very similar, it is reasonable to assume they have the same origin.  One possible experimental origin is a small inclusion of misaligned crystal in the sample.  Another possible origin is the complexity of the Fermi surface leads to different characteristic fields as recently observed in URu$_2$Si$_2$ \cite{Altarawneh2011}.

Resistivity measurements taken on the same setup are also shown in Figure 2.  In 0\,T, the resistivity shows a T$^2$ dependence between T$_{C}$ and T$_{SC}$ indicating a conventional Fermi liquid.  As the field is increased, the temperature dependence of the resistivity becomes less like a Fermi liquid in the measured temperature range which is expected as enhanced fluctuations drive the T$^2$ behavior to low temperature.  At H$^*$  the resistivity can be fitted with  $\rho\sim T^{\frac{5}{3}}$ from 1.5\,K to T$_{SC}$ in agreement with FM spin fluctuation theory close to a FM critical point \cite{Moriya1985}. This coupled with $\frac{S}{T}\sim\ln(T)$ at H$^*$ (see figure 1), which has been observed at a antiferromagnetic critical point \cite{Hartmann2010} but can be applied to weakly ferromagnetic metals \cite{Moriya1985}, stresses that H$^*$ is an FM critical point. Our experiment is only performed to 16\,T, i.e still close to H$^*$,  in higher magnetic fields a $\rho\sim T^2$ law must be recovered although this can be masked by quantum orbital effects when measuring magnetoresistivity in the transverse configuration.  The enhanced FM correlations at this point would lead to an enhanced H$_{c2}$ similar to the mechanism proposed \cite{Miyake2008} for reentrant superconductivity in URhGe.  However this scenario does not explain the presence of two anomalies in the thermopower.  This plus the evidence of a Fermi surface modification in high magnetic field implies that the Fermi surface is strongly affected by the applied field. Enhancement of superconductivity through topological changes of the Fermi surface has already been considered for CeCu$_2$Si$_2$ \cite{Thomas1996}. In the following we argue that the Fermi surface modification under field can lead to the enhanced T$_{SC}$ in UCoGe.

In a metal the thermopower can be written as
\begin{equation}
\frac{S}{T}=-\frac{\pi^2k_B^2}{3e}\left[\frac{1}{A_k}\frac{\delta A_k}{\delta E}+\frac{1}{l}\frac{\delta l}{\delta E}\right]_{E=E_f}
\end{equation}
where $A_k$ is the area of the Fermi surface in reciprocal space, $l$ is the mean free path and $E_f$ is the Fermi energy. For the following we assume $l$ is energy independent and therefore $S/T$ is dominated by the first term in the equation.  A large thermopower is observed when the area of the Fermi surface is strongly dependent on energy.  This is the case near an extremum in the band where a small change in Fermi energy will change the Fermi surface area by a large amount. At finite temperature the observed increase in thermopower would be broadened by a factor $kT$ as observed in the Figure 2(c).  If the Fermi energy is at an extremum in the band then the Fermi velocity $v_f$ is also very small ($v_f=\left|\frac{\delta E_k}{\delta k}\right|_{E_f}$).  A decreased Fermi velocity will increase the orbital limit of superconductivity and therefore increase $H_{c2}$ in the absence of Pauli limiting (as is the case in UCoGe). In UCoGe, if the anomaly in $S/T$ at $H^*$ is the result of the band structure being modified by the field to a point with a reduced $v_F$ then the orbital limit of superconductivity will increase and $H_{c2}$ will appear to increase.  Figure 2(a) shows an H$_{c2}$ curve calculated for a polar band state \cite{Scharnberg1980} with a $v_F$ assumed to be related to $S/T$ at low temperature ($S/T\sim1/v_F$ normalized to the zero field values) up to $H^*$.  $H^{**}$ would be a second band being modified and then at higher fields the hole like Fermi surface has been suppressed and an electron like Fermi surface is recovered.

To date, there has been one band structure calculation reported \cite{Samsel-Czekaa2010}. The calculation predicts there is a Fermi surface change between the ferromagnetic and paramagnetic states.  The paramagnetic state has three bands which cross the Fermi level and the ferromagnetic state has four.  Both states have a small Fermi surface which is roughly consistent with the measured thermopower.  In the ferromagnetic state, there are several extremum in bands close to the Fermi level particularly close to the $\Gamma$ point of the Brillouin zone which could result in an increased thermoelectric response.  However, to confirm this band structure, a full quantum oscillations study needs to be performed.  This is similar to UGe$_2$ where Fermi surface reconstruction and critical spin fluctuations are associated presumably due to the presence of flat bands near the Fermi level \cite{Flouqet2005}. Recently, similar conclusions have also been reported for URhGe \cite{Yelland2011}.

\section{Conclusion}

In conclusion, the zero field thermopower of UCoGe implies a low carrier metal in the ferromagnetic state.  Applying a field perpendicular to the easy axis, enhances the superconductivity and the thermopower exhibits a sharp peak at low temperature in proximity to an enhanced T$_{SC}$. The temperature dependence of the thermopower is consistent with ferromagnetic fluctuations being driven to 0\,K. At higher fields, a second anomaly is observed and the thermopower changes sign implying a Fermi surface reconstruction.  Clearly a Fermi surface change must also be included in theoretical descriptions.  Our results stress the importance of taking into the account the Fermi surface in future theoretical studies of ferromagnetism around a critical point and emphasize the necessity to have a detailed understanding of the electronic band structure.

We acknowledge the financial support of the French ANR within the programs DELICE, CORMAT, SINUS and the European Commission from the 7th framework programme 'Transnational access', contract No 228043-EuromagNETII-Integrated Activities. We thank K. Behnia for his comments.

\end{document}